\documentclass[aps]{revtex4}
\usepackage{amsmath,amssymb}
\usepackage[dvips]{graphicx}

\newcommand{\MeV}{\;\text{MeV}}
\newcommand{\Nc}{N_{\text{c}}}
\newcommand{\Nf}{N_{\text{f}}}
\newcommand{\muB}{\mu_{\text{B}}}
\newcommand{\muq}{\mu_{\text{q}}}
\newcommand{\sfree}{s_{\text{free}}}
\newcommand{\nfree}{n_{\text{free}}}
\newcommand{\U}{\mathcal{U}}
\newcommand{\bPhi}{\bar{\Phi}}

\begin{document}
\title{Effective Model Approach to the Dense State of QCD Matter}
\author{Kenji Fukushima}
\affiliation{Yukawa Institute for Theoretical Physics,
 Kyoto University, Kyoto 606-8502, Japan}
\begin{abstract}
 The first-principle approach to the dense state of QCD matter,
 i.e.\ the lattice-QCD simulation at finite baryon density, is not
 under theoretical control for the moment.  The effective model study
 based on QCD symmetries is a practical alternative.  However the
 model parameters that are fixed by hadronic properties in the vacuum
 may have unknown dependence on the baryon chemical potential.  We
 propose a new prescription to constrain the effective model
 parameters by the matching condition with the thermal Statistical
 Model.  In the transitional region where thermal quantities blow up
 in the Statistical Model, deconfined quarks and gluons should
 smoothly take over the relevant degrees of freedom from hadrons and
 resonances.  We use the Polyakov-loop coupled Nambu--Jona-Lasinio
 (PNJL) model as an effective description in the quark side and show
 how the matching condition is satisfied by a simple ans\"{a}tz on the
 Polyakov loop potential.  Our results favor a phase diagram with the
 chiral phase transition located at slightly higher temperature than
 deconfinement which stays close to the chemical freeze-out points.
\end{abstract}
\maketitle

%%%%%%%%%%   Introduction   %%%%%%%%%%

\section{Introduction}

Exploration of the QCD (Quantum Chromodynamics) phase diagram,
particularly toward a higher baryon-density regime, is of increasing
importance in both theoretical and experimental
sides \cite{Fukushima:2010bq}.  From the theoretical point of view, so
far, only the lattice-QCD simulation \cite{Fukushima:2010bq,%
DeTar:2009ef,Borsanyi:2010cj,Bazavov:2010sb} is the first-principle
calculation of QCD at work to explore the phase transitions associated
with chiral restoration and quark deconfinement.  The Polyakov loop
$\Phi$ and the chiral condensate $\langle\bar{\psi}\psi\rangle$ are
the (approximate) order parameters for quark deconfinement and chiral
restoration, respectively, which are gauge invariant and measurable on
the lattice (though both require renormalization corrections).  The
lattice-QCD simulation is, however, of limited practical use and it
works only when the baryon chemical potential $\muB$ is sufficiently
smaller than the temperature $T$.  For $\muB/T\gtrsim 1$ the notorious
sign problem prevents us from extracting any reliable information from
the lattice-QCD data \cite{Fukushima:2010bq,Muroya:2003qs}.

The effective model study is an alternative and pragmatic approach
toward the phase diagram of dense QCD.\ \ Some may complain that the
model study relies on not QCD directly but on just a model.  Results
from the model analysis are, nevertheless, what we can get at best for
the moment.  Even within the framework of the model study there are
several different attitudes.  One way for theorists to go is simplify
QCD so that it can be solvable without introducing further
approximations.  QCD-like models in lower dimensions (such as the
't~Hooft model) \cite{Schon:2000qy}, the strong-coupling expansion in
the lattice formulation \cite{Fukushima:2003vi}, and the large-$\Nc$
limit of QCD \cite{McLerran:2007qj} are typical examples in this
direction.

\begin{figure}
 \includegraphics[width=0.7\columnwidth]{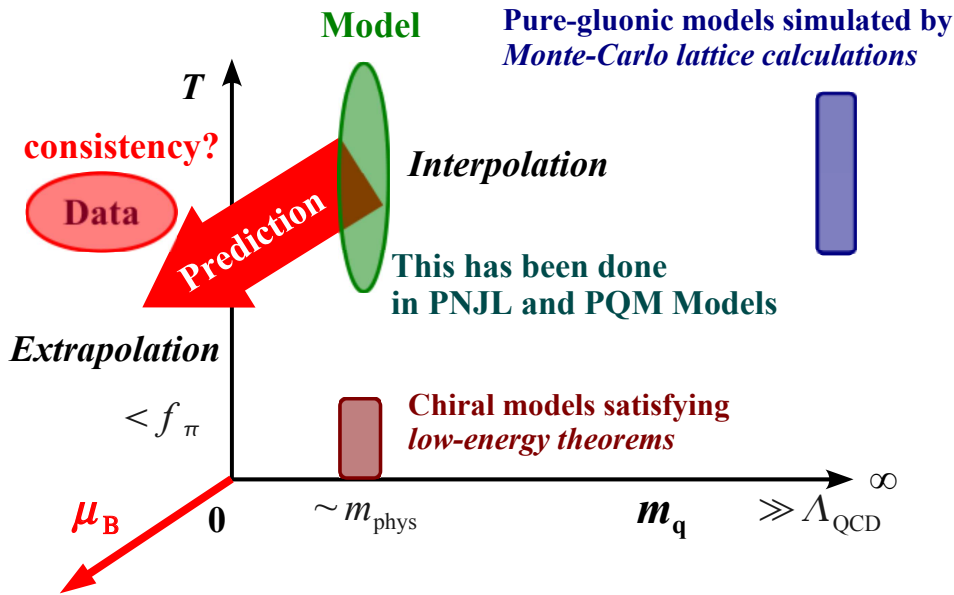}
 \caption{Schematic picture to show how the effective model is
   constructed as an interpolation between the pure-gluonic theory (at
   $m_q\ll\Lambda_{\text{QCD}}$ and $T\sim\Lambda_{\text{QCD}}$) and
   the chiral models (at $m_q\lesssim m_{\text{phys}}$ and
   $T<f_\pi$).  The prediction from the interpolated effective model
   is done as an extrapolation toward some new axis such as the baryon
   chemical potential, whose consistency with available data must be
   eventually checked.}
 \label{fig:schematic}
\end{figure}

Here, we shall take another way to proceed into the phase structure.
The idea is the following (as schematically illustrated in
Fig.~\ref{fig:schematic});
\begin{enumerate}
\item Construct a model that works for infinitely heavy quarks
  ($m_q\to\infty$) in such a way that the model respects the global
  symmetry (center symmetry) in the finite-$T$ pure gluonic sector.
\item Choose a chiral model based on the global symmetry (chiral
  symmetry) for massless quarks ($m_q\to 0$) in such a way that the
  spontaneous breakdown of chiral symmetry is correctly described.
\item Interpolate a finite-$m_q$ model between above-mentioned two.
  It is minimally required that the infinite $m_q$ limit and the
  vanishing $m_q$ limit should recover the above models respectively.
\item Check if the interpolation is properly chosen or not by
  comparing the model outputs to available lattice-QCD and/or
  phenomenological data.
\end{enumerate}

%---   figure   ---%

%---   figure   ---%

Along this line the model is not necessarily solvable and usually
needs some additional approximations.  Nevertheless, if the item $4$
above is taken into account very carefully, one may claim that one is
dealing with a phase diagram of QCD, not of QCD-like models, in a
sense that the situation one is handling is not $(1+1)$ dimensions,
not $g^2\to\infty$, and not $\Nc\to\infty$.  Sometimes, to this aim
toward the QCD phase diagram, one has to face ``dirty'' businesses;
it is often the case that the phase structure might be significantly
changed by uncontrollable model parameters, which one can take in
twofold ways --- pessimists would be disappointed and say that the
model cannot predict anything, and on the other hand, optimists would
be delighted and say that the model has clarified a non-trivial role
played by the model parameter in understanding the phase diagram.

Let us briefly explain how to implement the items $1$ and $2$.  In the
absence of particles transforming in the color fundamental
representation, the genuine gauge symmetry possessed by the pure
gluonic theory is $\mathrm{SU}(\Nc)/\mathrm{Z}(\Nc)$.  If one performs
the $\mathrm{Z}(\Nc)$ transformation on the gauge links, the fields
are shifted typically by $2\pi/(\Nc a)$ where $a$ is a characteristic
scale (lattice spacing).  The perturbation theory breaks this
$\mathrm{Z}(\Nc)$ symmetry but this is practically no problem because
the shift goes infinity as $a\to0$.  Furthermore one can generalize
the similar procedure onto not the individual gauge link but a product
of the $N_\tau$ gauge links along a finite extent.  Then, the fields
are shifted by $2\pi/(\Nc N_\tau a)=2\pi T/\Nc$, which remains
sensible in the $a\to0$ limit.  In this way the Polyakov loop matrix
$L$ is defined, that is, $L=\prod_\tau U_\tau$ and the
$\mathrm{Z}(\Nc)$ symmetry with respect to $L$ is called
``center symmetry'' which breaks in the perturbation theory
\cite{Svetitsky:1985ye}.

The expectation value of the traced Polyakov loop,
$\Phi=\langle\text{tr}L\rangle$, is the order parameter for the quark
deconfinement phase transition in the pure gluonic system.  The most
intuitive way to understand this comes from the property that $\Phi$
is related to the free energy gain of a static single quark placed in
a hot gluonic medium as $\Phi=\exp[-f_q /T]$.  Therefore $\Phi=0$
implies $f_q=\infty$, meaning that quarks never show up (confinement).
Once $\Phi$ becomes non-vanishing, $f_q$ should take a finite value
and thermal quark excitations are permitted.  The effective action for
$L$ or $\Phi$ has been computed perturbatively \cite{Gross:1980br} but
in order to discuss the phase transition from center symmetric to
center broken phases, one needs a non-perturbative evaluation of the
effective action.

Concerning the chiral dynamics, the model choice could be anything as
long as it can correctly describe the dynamical chiral symmetry
breaking pattern.  Then, the chiral properties are almost
automatically derived from the so-called low-energy (soft-pion)
theorems.  Of course some details of the phase transition such as the
critical temperature and the thermodynamic quantities depend on the
choice of the chiral model.  Because we are interested in the phase
transition associated with restoration of chiral symmetry, the
non-linear representation is inappropriate which is based on the
symmetry breaking.

The order parameter for the chiral phase transition is given by the
chiral condensate $\langle\bar{\psi}\psi\rangle$.  This is simple to
understand --- the quark mass $m_q$ and an operator $\bar{\psi}\psi$
are conjugate to each other, so $m_q$ is a source to generate
$\langle\bar{\psi}\psi\rangle$ and in turn
$\langle\bar{\psi}\psi\rangle$ is a source to generate the dynamical
mass that breaks chiral symmetry.  There are well-established chiral
models such as the Nambu--Jona-Lasinio (NJL) model and the quark-meson
(QM) model.

The interpolation at the item $3$ is the main problem.  There is no
theoretical justification at all for the existence of reasonable
interpolation.  We can judge how good or how bad it is only through
the comparison at the item $4$.  At this point it is already obvious
that the Polyakov loop model is not sufficient to access the realistic
QCD phase transition, though it may capture interesting
phenomenological consequences \cite{Dumitru:2000in}.  In a similar
sense conventional chiral models are not good enough to draw the QCD
phase diagram even though they are usually designed to be a good
description of hadronic properties in the vacuum
\cite{Hatsuda:1994pi}.  To address the QCD phase transitions the first
test for the validity of the model description should be the
consistency check with the known properties available from the
lattice-QCD simulation at $T\neq0$ and $\muB=0$;  the finite-$T$
behavior of two order parameters, $\Phi$ and
$\langle\bar{\psi}\psi\rangle$, and the thermodynamic quantities such
as the pressure, the internal energy density, the entropy density,
etc.

Along this line the Polyakov-loop coupled chiral models such as the
PNJL (Polyakov--NJL) \cite{Fukushima:2003fw,Ratti:2005jh} and the PQM
(Polyakov-QM) \cite{Schaefer:2007pw,Schaefer:2009ui} models are quite
successful to treat both order parameters on the equal footing.
Besides, the Polyakov loop potential $\U[\Phi]$ is determined by the
lattice data in the pure gluonic theory, namely, by the Polyakov loop
$\Phi(T)$ and the pressure $p(T)$ as functions of $T$.  This means
that the PNJL and PQM models include the pressure contribution from
gluons as well as quarks, so that the models are able to deal with the
full thermodynamics which are to be compared with the full lattice-QCD
simulation.  The important point is that the dynamics of transverse
gluons $A_i^T$ is also under the control of the deconfinement order
parameter $\Phi$ and thus is to be encompassed in the parametrization
of the Polyakov loop potential $\U[\Phi]$, while the Polyakov loop
itself is expressed in terms of the longitudinal gluon $A_4$.

Here we would like to emphasize that the success of the PNJL and PQM
models is far beyond the fitting physics.  Model parameters are fixed
separately in two regions, i.e.\ in the pure-gluonic theory (at
$m_q\ll\Lambda_{\text{QCD}}$ and $T\sim\Lambda_{\text{QCD}}$) and in
the chiral models (at $m_q\lesssim m_{\text{phys}}$ and $T<f_\pi$).
The interpolation procedure does not involve any further fitting.  It
is a highly non-trivial discovery that there exists a reasonable way
to make an interpolation fairly consistent with the full lattice-QCD
data.

The next step one should think of is the prediction from the model.
This is done by an extrapolation of the model toward some new axis
such as the baryon chemical potential.  By now there are different
versions of the ``QCD phase diagram'' drawn in this way by means of
the PNJL and PQM models with different parameter tunings
\cite{Fukushima:2003fw,Ratti:2005jh,Schaefer:2007pw,Schaefer:2009ui,%
Sasaki:2006ww}.  If we go into small details, there are many places
where we can talk about the difference.  Here we shall limit ourselves
to look at the difference mainly in the behavior of the Polyakov loop
or the deconfinement (crossover) transition line.  Some of the model
results may be close to the true answer, and some may not.  We must
have a guiding principle to select out which is preferred and which is
not.  The available and reliable data at finite baryon density is,
however, extremely limited.  In what follows we shall elucidate the
idea and find that the naive extrapolation from the PNJL and PQM
models is not acceptable.  To see this, we will explain the results
from the thermal Statistical Model in the next section.

%%%%%%%%%%   Thermodynamics from the Statistical Model   %%%%%%%%%%

\section{Thermodynamics from the Statistical Model}

%---   figure   ---%

\begin{figure}
 \includegraphics[width=0.7\columnwidth]{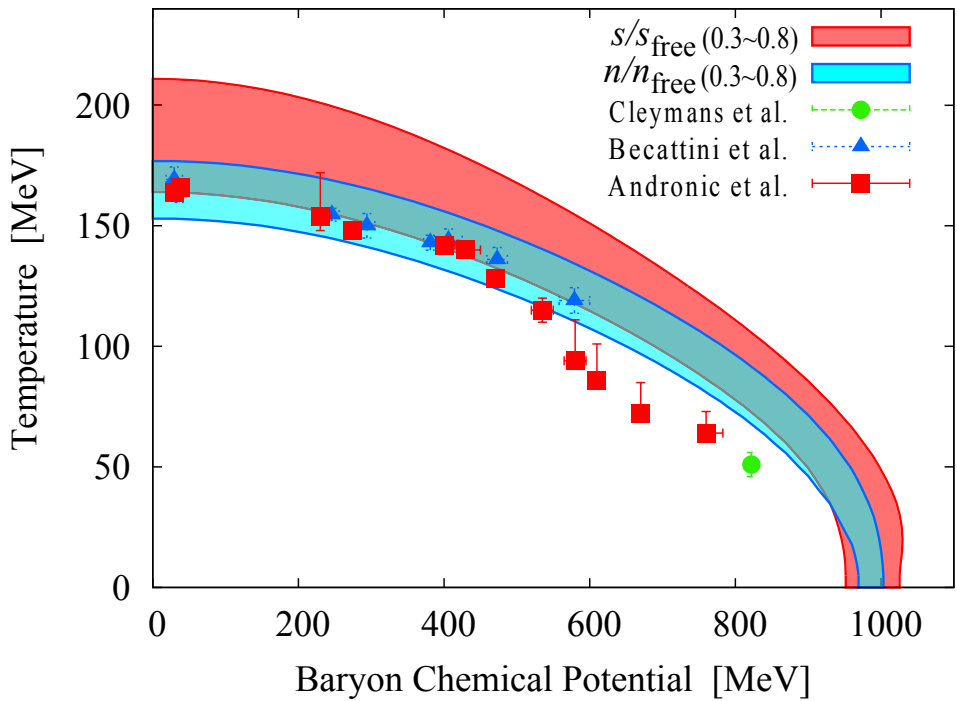}
 \caption{Chemical freeze-out points taken from
   Refs.~\cite{Cleymans:2005xv,Becattini:2005xt}.  The red and blue
   (upper and lower) bands represent the regions where the entropy
   density $s$ and the baryon number density $n$, respectively,
   increase quickly from $0.3$ to $0.8$ in the unit
   of free quark-gluon values, $\sfree$ and $\nfree$ (see
   Eq.~\eqref{eq:free}).}
 \label{fig:freezeout}
\end{figure}

%---   figure   ---%

Regarding the QCD phase diagram at finite $T$ and $\muB$ useful
information is quite limited.  The lattice-QCD at finite density is
being improved, but still different techniques to circumvent the sign
problem lead to different results.  Only the chemical freeze-out
points in the heavy-ion collisions are experimental hints about the
phase diagram.  Although the freeze-out points shape an intriguing
curve on the $\muB$-$T$ plane, as plotted by error-bar dots in
Fig.~\ref{fig:freezeout}, one should carefully interpret it.

The freeze-out points are not the raw experimental data but
\textit{an} interpretation through the thermal Statistical
Model \cite{Cleymans:2005xv,Becattini:2005xt}.  In this model the
grand canonical partition function is given by contributions from the
non-interacting gas of hadrons and resonances.  In view of the fact
that the Statistical Model is such successful to fit various particle
ratios with $\muB$ and $T$ only ($\mu_Q$, $\mu_{\text{s}}$, and
$\mu_{\text{c}}$ are determined by the collision condition), it should
be legitimate to take the freeze-out points for \textit{experimental}
data, which in turn validates the Statistical Model (though why it
works lacks for an explanation from QCD).

Let us proceed by further accepting that the Statistical Model is a
valid description of the state of matter until the freeze-out curve or
even slightly above.  It is then a straightforward application of the
Statistical Model to estimate thermodynamic quantities such as the
pressure $p$, the entropy density $s$, the baryon number density $n$,
etc.  We here utilize the open code THERMUS ver.2.1 to calculate $s$
and $n$ at various $T$ and $\muB$ \cite{Wheaton:2004qb}.  From now on
the Statistical Model analysis specifically means the use of THERMUS.

Figure~\ref{fig:freezeout} shows the chemical freeze-out points taken
from Refs.~\cite{Cleymans:2005xv,Becattini:2005xt}, on which $s$ and
$n$ are overlaid.  For convenience we normalized these quantities by
\begin{equation}
 \begin{split}
 \sfree & = \biggl\{ (\Nc^2-1) + \frac{7}{4} \Nc\Nf\biggr\}
  \frac{4\pi^2}{45}T^3 + \frac{\Nc\Nf}{3} \muq^2 T \;,\\
 \nfree &= \Nf \biggl( \frac{\muq^3}{3\pi^2}
  + \frac{\muq T^2}{3} \biggr) \;.
 \end{split}
\label{eq:free}
\end{equation}
These are the entropy density and the baryon number density of free
massless $\Nc^2-1$ gluons and $\Nc\Nf$ quarks.

Here we note that, in drawing Fig.~\ref{fig:freezeout}, we have
intentionally relaxed the neutrality conditions for electric charge
and heavy flavors and simply set
$\mu_Q=\mu_{\text{s}}=\mu_{\text{c}}=0$.  We have done so in order to
make it possible to compare the results from the Statistical Model to
the chiral effective model in later discussions.  [We note that one
  can force the chiral model to satisfy neutrality but it would be
  technically involved \cite{Fukushima:2009dx}.]  Nevertheless, we
would emphasize that the neutrality conditions have only minor effects
on the bulk thermodynamics and make only small differences in any
case.  We should also mention that we used Eq.~\eqref{eq:free} with
$\Nc=\Nf=3$. The choice of $\sfree$ and $\nfree$ (and relevant $\Nf$)
is arbitrary and the following discussions do not rely on this
particular choice, for we will use $\sfree$ and $\nfree$ just as
common denominators to display the Statistical Model results and the
PNJL model results.

The Statistical Model cannot tell us about the QCD phase transitions.
Still, Fig.~\ref{fig:freezeout} is already suggestive enough.  We can
clearly see the thermodynamic quantities from the Statistical Model
blowing up in a relatively narrow region.  The red and blue (upper and
lower) bands indicate the regions where $s/\sfree$ and $n/\nfree$,
respectively, grow quickly from $0.3$ to $0.8$.  In the Hagedorn's
picture \cite{Cabibbo:1975ig} this rapid and simultaneous rise in $s$
and $n$ has a natural interpretation as the Hagedorn limiting
temperature above which color degrees of freedom is liberated,
i.e.\ color deconfinement.

The idea here is to make use of the thermodynamics from the
Statistical Model as shown in Fig.~\ref{fig:freezeout} to judge if the
Polyakov-loop coupled chiral models work fine at finite density.  We
also make an important remark that this idea can be effective only up
to about $\muB\lesssim 400\sim 600\MeV$ because the chemical
freeze-out points start dropping down steeply in this density region,
which suggests an onset of some new form of matter;  an example of
such possibilities is quarkyonic matter \cite{Andronic:2009gj}.

%%%%%%%%%%   Thermodynamics from the PNJL Model   %%%%%%%%%%

\section{Thermodynamics from the PNJL Model}

Figure~\ref{fig:freezeout} is useful to make a guesstimate about the
deconfinement boundary but we can deduce no information about the
chiral property.  This is because the thermal Statistical Model needs
no medium modification driven by chiral restoration.  So, to address
the QCD phase transitions and associated boundaries, we must find a
way to connect the thermodynamics in Fig.~\ref{fig:freezeout} to the
order parameters $\Phi$ and $\langle\bar{\psi}\psi\rangle$.  Here let
us go into details of the chiral effective model.

It is crucial to adopt the Polyakov-loop coupled model here because
the entropy density should contain contributions from gluons which are
taken care of by the Polyakov loop potential $\U[\Phi]$.  The PNJL
model that we use below is defined with the following potential;
\begin{equation}
 \U[\Phi,\bPhi] = T^4 \biggl\{ -\frac{a(T)}{2}\bPhi\Phi
  + b(T)\ln\Bigl[ 1-6\bPhi\Phi+4(\bPhi^3+\Phi^3)
  -3(\bPhi\Phi)^2 \Bigr] \biggr\}
\end{equation}
with $a(T)=a_0 + a_1(T_0/T) + a_2(T_0/T)^2$ and $b(T)=b_3(T_0/T)^3$.
There are thus five parameters one out of which is constrained by the
Stefan-Boltzmann limit, i.e.\ $\U\to -(8\pi^2/45)T^4$ at
$\Phi=\bPhi=1$ in the $T\to\infty$ limit.  These parameters are fixed
by the pure-gluonic lattice data as $a_0=3.51$, $a_1=-2.47$,
$a_2=15.2$, $b_3=-1.75$ \cite{Ratti:2005jh}, and $T_0=270\MeV$ from
the deconfinement temperature of first order in the pure-gluonic
theory \cite{Boyd:1996bx}.  It is important to note that only $T_0$ is
a dimensional parameter, so that the energy scale is set by this
$T_0$ alone.

In addition the NJL sector of the PNJL model has five more
parameters in the three-flavor case \cite{Fukushima:2003fw} appearing
in the mean-field thermodynamic potential;
\begin{equation}
 \begin{split}
 \Omega_{\text{NJL}} & = g_S\bigl( \langle\bar{u}u\rangle^2
  +\langle\bar{d}d\rangle^2 + \langle\bar{s}s\rangle^2 \bigr)
  +4g_D \langle\bar{u}u\rangle \langle\bar{d}d\rangle
  \langle\bar{s}s\rangle
  -2\Nc\sum_i \int^\Lambda \!\frac{d^3 p}{(2\pi)^3}
  \:\varepsilon_i(p) \\
 &\quad -2T\sum_i \int\frac{d^3 p}{(2\pi)^3} \biggl\{ \ln\det
  \biggl[ 1+ L\, e^{-(\varepsilon_i(p)-\muq)/T} \biggr] + \ln\det
  \biggl[ 1+ L^\dagger\, e^{-(\varepsilon_i(p)+\muq)/T} \biggr]
  \biggr\} \;,
 \end{split}
\end{equation}
where the energy dispersion relation $\varepsilon_i(p)$ depends on the
flavor index $i$ as $\varepsilon_i(p)^2 = p^2+M_i^2$ and the
constituent quark masses are
$M_u=m_u -2g_S\langle\bar{u}u\rangle -2g_D\langle\bar{d}d\rangle
\langle\bar{s}s\rangle$ and so on.  The model parameters are then the
light and heavy quark masses $m_{\rm ud}$ and $m_{\rm s}$, the momentum
cutoff $\Lambda$, the four-fermionic interaction strength $g_S$, and
the $\mathrm{U(1)_A}$-breaking six-fermionic interaction strength
$g_D$, which are all fixed by the pion mass $m_\pi$, the kaon mass
$m_K$, the eta-prime mass $m_{\eta'}$, the pion decay constant
$f_\pi$, and the chiral condensate $\langle\bar{\psi}\psi\rangle$
\cite{Hatsuda:1994pi}.

%---   figure   ---%

\begin{figure}
 \includegraphics[width=0.7\columnwidth]{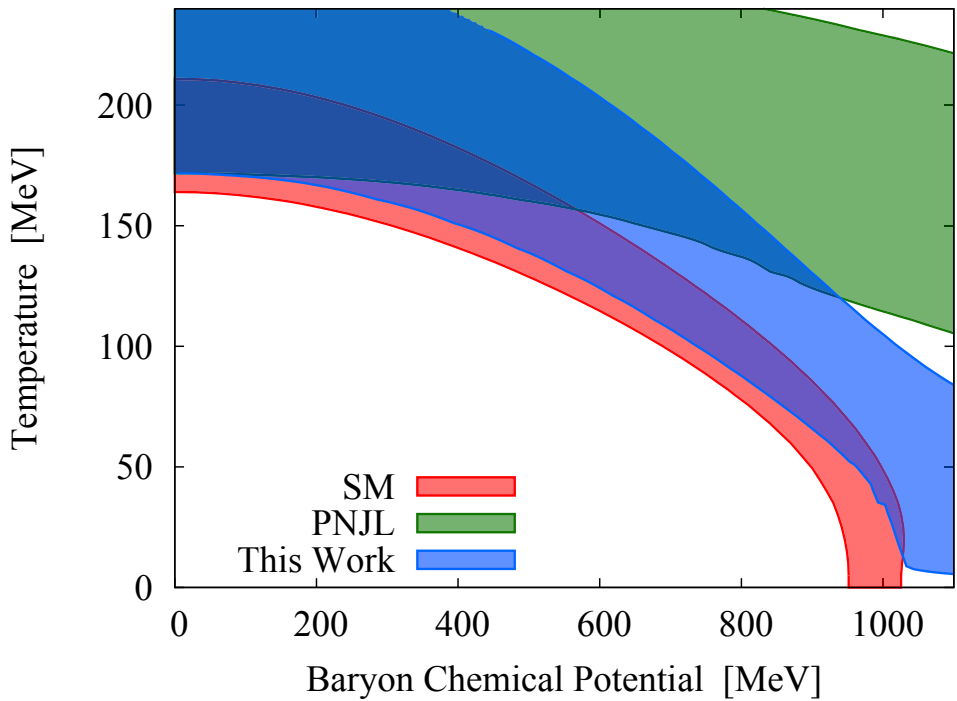}
 \caption{Entropy density normalized by $\sfree$ (from $0.3$ to $0.8$)
   in the Statistical Model (bottom band with red color; same as shown
   in Fig.~\ref{fig:freezeout}) and that in the PNJL model with a
   choice $T_0=200\MeV$ (top band with green color).  The blue band
   between two represents the results with the
   ans\"{a}tz~\eqref{eq:T0}.}
 \label{fig:entropy}
\end{figure}

%---   figure   ---%

In the presence of dynamical quarks, if we keep using the pure-gluonic
critical temperature $T_0=270\MeV$, the simultaneous crossover
temperature of deconfinement and chiral restoration becomes over
$200\MeV$, which is too high as compared to the lattice-QCD value.  It
is nicely argued in Ref.~\cite{Schaefer:2007pw} that the back-reaction
from quark loops affects the mass scale $T_0$ which changes from
$T_0=270\MeV$ for $\Nf=0$ down to $T_0=208\MeV$ for $\Nf=2$ and
$T_0=187\MeV$ for $\Nf=2+1$ \cite{Schaefer:2007pw}.  Here we choose to
use $T_0=200\MeV$ for calculations at $\muB=0$ throughout.

In Fig.~\ref{fig:entropy} we show the entropy density calculated in
the mean-field PNJL model with $T_0=200\MeV$ fixed, in the same way as
in the Statistical Model drawn in Fig.~\ref{fig:freezeout}.  The
bottom (top) band in red (green) color is the result from the
Statistical Model (PNJL model).  From the figure it is obvious that
the PNJL model is not consistent with the Statistical Model even at
the qualitative level.  With the properly scaled $T_0$ from $270\MeV$
down to $200\MeV$, the blow-up behavior in $s$ from the Statistical
Model can be smoothly connected to the PNJL model description only in
the region up to $\muB\lesssim 400\MeV$.  The curvature of the band as
a function of $\muB$ is significantly different;  the PNJL model
result is too flat horizontally and the green band eventually takes
apart from the red region where the Statistical Model breaks down.

%%%%%%%%%%   Matching Prescription   %%%%%%%%%%

\section{Matching Prescription}

Such a manifest discrepancy from the Statistical Model is a critical
drawback of the PNJL model.  The situation is not changed even in the
PQM model as long as $T_0$ is a constant.  To make the entropy density
at $\muB\gtrsim 400\MeV$ get saturated earlier as is the case in the
Statistical Model, quark degrees of freedom must be released at
smaller temperature than predicted by the PNJL model.

One can imagine how this drawback occurs in the PNJL model study;  the
energy scale in the Polyakov loop potential is specified by the
dimensional parameter $T_0$ that may differ depending on the effects
of $T$ and $\muB$ in the quark sector.  We have shifted $T_0$ from
$270\MeV$ down to $200\MeV$, through which we have incorporated the
scale change induced by $\Nf$ quarks at finite $T$.  In this way we
may well consider that $T_0$ should decrease with increasing
$\muB$ as pointed out in Ref.~\cite{Schaefer:2007pw}.

Our idea proposed here is to make use of Fig.~\ref{fig:entropy} to fix
$T_0(\muB)$ for consistency with phenomenology.  One can pick up other
thermodynamic quantities than the entropy density like the internal
energy density, which would anyway make little change in the final
result.  Besides, the choice of the entropy density is most natural
because it counts the effective degrees of freedom and thus is a
sensitive quantity to probe deconfinement.  In
Ref.~\cite{Cleymans:2005xv} the freeze-out curve is parametrized as
\begin{equation}
 T_{\rm f}(\muB) = a - b\,\muB^2 - c\,\muB^4
\end{equation}
with the fitting result $a=166(2)\MeV$,
$b=1.39(16)\times10^{-4}\MeV^{-1}$, and
$c=5.3(21)\times10^{-11}\MeV^{-3}$.  Because the behavior of the
entropy density must be dominantly controlled by deconfinement, we
postulate that the change in $T_0$ is to be correlated with
$T_{\rm f}(\muB)$.  We see that the freeze-out points and the entropy
band in Fig.~\ref{fig:freezeout} have roughly same curvature indeed.
Let us simply use same $b$ and make an ans\"{a}tz as
\begin{equation}
 \frac{T_0(\muB)}{T_0} = 1 - (b T_0)
  \biggl(\frac{\muB}{T_0}\biggr)^2
  = 1 - 2.78\times 10^{-2} \biggl(\frac{\muB}{T_0}\biggr)^2 \;,
 \label{eq:T0}
\end{equation}
which yields the blue band in the middle of Fig.~\ref{fig:entropy}.
[To prevent unphysical negative $T_0$ for large $\muB$ we set a
  threshold at $10\MeV$ so that $T_0 \geq 10\MeV$.  Hence, the results
  at $T<10\MeV$ are not meaningful.]  We see at a glance that the
results from this modified PNJL model have a reasonable overlap with
the Statistical Model results in the whole density region as plotted.

At this point one might have thought of several questions.  First, the
ans\"{a}tz~\eqref{eq:T0} might look ad hoc, but we point out that our
choice happens to be very close to the independent argument in
Ref.~\cite{Schaefer:2007pw}, in which the $\muB$-dependence has been
estimated from the running coupling constant as
$T_0(\muB)=T_\tau e^{-1/(\alpha_0 b(\muB))}$ which is expanded to be
$T_0(\muB)/T_0 \simeq 1-2.1\times 10^{-2}(\muB/T_0)^2 + \cdots$.  In
the perturbative manner one can also understand how the
$\muB$-dependence enters the Polyakov loop potential which consists of
the closed loop of dressed gluon propagator.  The quark--anti-quark
polarization diagrams inserted in the gluon propagator generate the
back-reaction dependent on $\muB$.  There is another phenomenological
ans\"{a}tz for the $\muB$-dependent $\U[\Phi]$
\cite{Dexheimer:2009hi}.  Second, one might wonder if the energy scale
in the quark (NJL) sector should be modified as well.  Such
modification is not necessary, however.  This is because, as we have
mentioned, the Statistical Model requires no modification associated
with the chiral dynamics, which strongly implies that we do not have
to introduce $\muB$ dependent changes in the NJL parameters.  Third,
the ans\"{a}tz~\eqref{eq:T0} has terms only up to the quadratic order.
This means that we cannot go to high-density regions
with $\muB\ll T_0$.  This is indeed so and we have actually truncated
higher-order terms in Eq.~\eqref{eq:T0}.  In any case, as we have
noted before, the idea of the entropy matching holds only up to
$\muB\lesssim 400\sim 600\MeV$, and we should not take the results in
the high-density region seriously.

%%%%%%%%%%   Phase Diagram   %%%%%%%%%%

\section{Phase Diagram}

Now we get ready to proceed to the possible QCD phase diagram that is
fully consistent with the Statistical Model thermodynamics in
Fig.~\ref{fig:freezeout}.  Using the standard computational procedure
of the mean-field PNJL model we can solve $\Phi$ and
$\langle\bar{\psi}\psi\rangle$ as functions of $T$ and
$\muB$, from which the phase boundaries of deconfinement and chiral
restoration are located.

%---   figure   ---%

\begin{figure}
 \includegraphics[width=0.7\columnwidth]{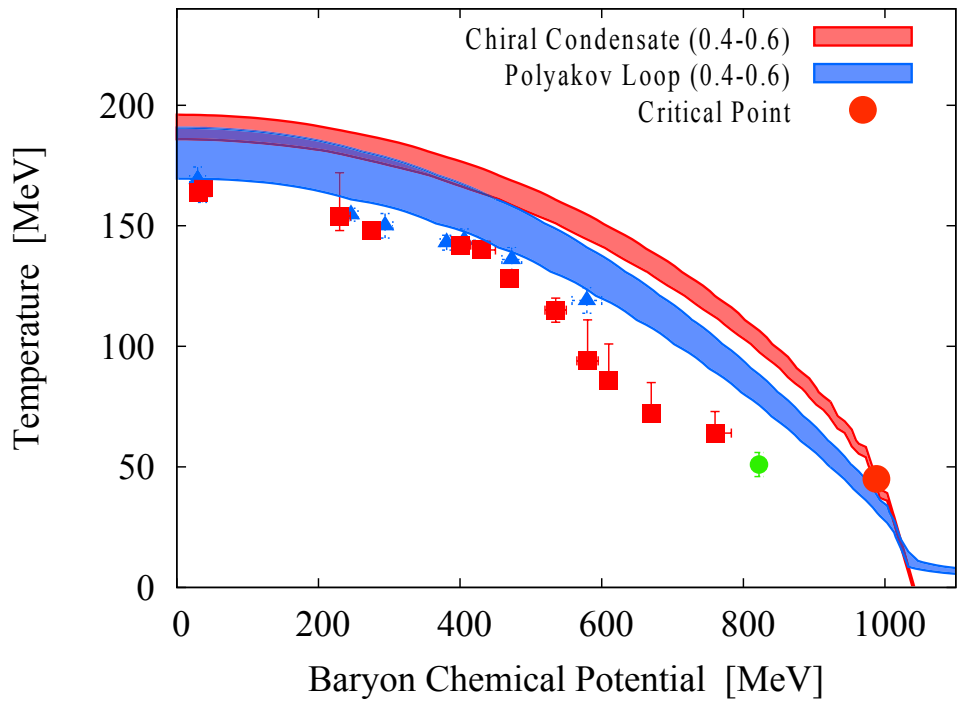}
 \caption{A figure taken from Ref.~\cite{Fukushima:2010is}.  Phase
   boundaries associated with deconfinement (blue band) and chiral
   restoration (red band).  Each band represents a region where the
   (normalized) order parameter develops from $0.4$ to $0.6$.}
 \label{fig:diagram}
\end{figure}

%---   figure   ---%

Figure~\ref{fig:diagram} (the central result of
Ref.~\cite{Fukushima:2010is}) shows the phase diagram from the
modified PNJL model.  The blue (red) band is a region where the
Polyakov loop (normalized light-quark chiral condensate) increases
from $0.4$ to $0.6$.  In contrast to the old PNJL model, the new
results indicate that the chiral phase transition is almost parallel
to and entirely above the deconfinement, which agrees with the
situation considered phenomenologically in
Ref.~\cite{Castorina:2010gy}.  We have found the critical point
\cite{Asakawa:1989bq,Stephanov:1998dy} at
$(\muB,T)\simeq(45\MeV,330\MeV)$, but should not take it seriously
since its location is beyond the validity region of the current
prescription.

%%%%%%%%%%   Conclusions   %%%%%%%%%%

\section{Conclusions}

It is an intriguing observation that the chiral phase transition
occurs later than deconfinement.  This is quite consistent with the
Statistical Model assumption.  In the Statistical Model the hadron
masses are just the vacuum values and any hadron mass/width
modifications are not considered, which would be a reasonable
treatment only if the chiral phase transition is separated above the
Hagedorn temperature.  Under such a phase structure, besides, our
assumption of neglecting $\muB$-dependence in the NJL-model parameters
turns out to be acceptable in a similar sense as the Statistical Model
treatment.  This can be understood from the fact that the NJL part
yields the hadron masses in the vacuum which are intact in the
Statistical Model.

The failure of the standard PNJL model is attributed to baryonic
degrees of freedom which is missing;  the singlet-part of the thermal
excitation in the PNJL model can be translated into an expression in
terms of baryons as
\begin{equation}
 \int \frac{d^3 k}{(2\pi)^3}\; e^{-\Nc (\sqrt{k^2+M_q^2} - \muq)/T}
  = \frac{1}{\Nc^3}\int \frac{d^3 k'}{(2\pi)^3}\;
  e^{-(\sqrt{{k'}^2 + M_N^2} - \muB)/T} \;,
\end{equation}
where $M_N=\Nc M_q$, $\muB=\Nc \muq$, and $k'=\Nc k$ with which a
factor $1/\Nc^3$ appears from the integration measure.  Therefore, the
PNJL model significantly underestimates the baryonic excitations by
$1/\Nc^3$.  Hence, one may say that a modification made in $\U[\Phi]$
by hand stems, in principle, from confinement effects, which can be
presumably parametrized by the Polyakov loop.  This idea is
reminiscent of the treatment of transverse gluons.  It is an important
question how our phenomenological input \eqref{eq:T0} is
validated/invalidated from the first-principle QCD calculation.  This
may be answered by future developments in the functional
renormalization group method \cite{Braun:2009gm}.

%%%%%%%%%%   References   %%%%%%%%%%

\end{document}